\documentclass[12pt]{iopart}

\expandafter\let\csname equation*\endcsname\relax
\expandafter\let\csname endequation*\endcsname\relax
\usepackage{amsmath}

\usepackage{graphicx}
\usepackage{dcolumn}
\usepackage{bm}

\usepackage[utf8]{inputenc}
\usepackage[T1]{fontenc}

\usepackage{mathptmx}

\begin{document}

\title{Emission of electromagnetic radiation due to spin-flip transitions in a ferromagnet}

\author{E.A. Karashtin}
 \address{Institute for physics of microstructure of RAS, Nizhniy Novgorod, 603950, GSP-105, Russia}
 \address{Lobachevsky University, Gagarin ave., 23, Nizhniy Novgorod, 603950, Russia}
 \ead{eugenk@ipmras.ru}

\begin{abstract}

We theoretically analyze a possibility of electromagnetic wave emission due to electron transitions between spin subbands in a ferromagnet. Different mechanisms of such spin-flip transitions are cousidered. One mechanism is the electron transitions caused by magnetic field of the wave. Another mechanism is due to Rashba spin-orbit interaction. While two mentioned mechanisms exist in a homogeneously magnetized ferromagnet, there are two other mechanisms that exist in non-collinearly magnetized medium. First mechanism is known and is due to the dependence of exchange interaction constant on the quasimomentum of conduction electrons. Second one exists in any non-collinearly magnetized medium. We study these mechanisms in a non-collinear ferromagnet with helicoidal magnetization distribution. The  estimations of probabilities of electron transitions due to different mechanisms are made for realistic parameters, and we compare the mechanisms. We also estimate the radiation power and threshold current in a simple model in which spin is injected into the ferromagnet by a spin-polarized electric current through a tunnel barrier.

\end{abstract}

\maketitle

\section{\label{Intro}Introduction}
Electron transitions between energy bands may be accompanied by electromagnetic wave generation. This paper is devoted to the mentioned phenomenon when electrons jump between spin subbands in a ferromagnet. In a simple conductor without a spin-orbit interaction or applied magnetic field the energy states of conduction electrons are doubly degenerate with respect to spin. In a homogeneous ferromagnet their spin states are split into two subbands. The energy gap between these subbands usually corresponds to infrared or terahertz frequency, depending on the material \cite{PhysQuan, Weber}. Therefore study of electron transitions between the spin subbands with electromagnetic wave generation is important. This is encouraged by a possibility of precise control of magnetic state of the ferromagnet by applying magnetic field or fabrication of different nanostructures \cite{Udalov12, Fraerman1, Demidov, Mironov, Sapozhnikov1, Sapozhnikov2, Sapozhnikov3}. 

Terahertz wave generation by ferromagnets which are irradiated by a femtosecond optical pulse was developed recently \cite{Kampfrath, Seifert, Bulgarevich, Adam, Nandi}. Usually this effect is explained by intraband electron transitions, i.e. processes without spin-flip processes \cite{Aktsipetrov}. The terahertz frequency range is due to the pulse length (typically it is $10-50fs$ which corresponds to $20-100THz$). The emission spectrum here is characterized by broad frequency range. Considering the interband electron transitions, i.e. transitions with spin flip, the frequency range is determined by the energy gap between spin subbands. Besides, stimulated emisson of electromagnetic wave may be obtained in such case.

It is well known that electro-dipole transitions of electrons between spin subbands are forbidden in a homogeneous ferromagnet. In order to show this, let us consider the Vonsovsky s-d exchange model \cite{Vonsovsky}. In this model, the magnetization $\bm{M}$ is created by localized d- or f-electrons, and the conduction electrons which are supposed to be s-electrons have the following hamiltonian:
\begin{equation} \label{Eq1_ham}
\hat{H} = \frac{\hat{\bm{p}}^2}{2 m_e} \Delta + J \left(\hat{\bm{\sigma}} \cdot \bm{M}\right),
\end{equation}
where $\hat{\bm{p}} = -i \hbar \nabla$ is the electron momentum operator, $m_e$ is the electron mass, $J$ is the exchange constant, $\hat{\bm{\sigma}}$ is vector of Pauli matrices which is proportional to the operator of electron spin. Here $\bm{M}$ is supposed to be a constant vector along the Cartesian z-axis. In equilibrium, the electron wavefunctions are 
\begin{equation} \label{Eq1_wf}
\psi_+ = exp\left(i \bm{k r}\right)
\begin{pmatrix}
1  \\
0
\end{pmatrix},
\psi_- = exp\left(i \bm{k r}\right)
\begin{pmatrix}
0  \\
1
\end{pmatrix},
\end{equation}
where $\bm{k}$ is the electron wave vector, $\bm{r}$ is the vector of Cartesian coordinates. The wave functions of two spin subbands correspond to electrons with average spin either parallel or antiparallel to $\bm{M}$. These wavefunctions are orthogonal to each other. In the electro-dipole approximation, only the electric field of the wave $\bm{E}_\omega = \bm{E} exp\left(-i \omega t\right)$ is taken into account ($\omega$ is the wave frequency). The coordinate dependence of this electric field leads to spatial dispersion and thus should be neglected here. In order to study the interaction of electrons with an electromagnetic wave in such approximation, it is convenient to take the gauge at which the electric potential is zero. Thus the electron momentum operator $\hat{\bm{p}}$ should be changed to $\left(\hat{\bm{p}} - \frac{e}{c} \bm{A}_\omega\right)$ \cite{Landau3, Landau8}, where $\bm{A}_\omega = -i\frac{c}{\omega} \bm{E}_\omega$ is the vector-potential. The operator of interaction of electrons and electromagnetic wave takes the form
\begin{equation} \label{Eq1_em}
\hat{H}_{e-m} = \frac{e \hbar}{2 m_e \omega} \left(\bm{E}_\omega \cdot \nabla\right),
\end{equation}
where $e$ is the absolute electron charge. This interaction is called the minimal coupling \cite{Tatara}. It is obvious that since the operator (\ref{Eq1_em}) does not change the electron spin its matrix element is zero and thus the electron transition probability due to interaction with electromagnetic wave is zero.

This situation can be changed by taking into account some additional interaction or condition that breaks such symmetry. It was predicted earlier that if the exchange constant depends on the electron momentum $J=J\left(\hat{\bm{p}}\right)$ the electron transitions are allowed \cite{KadigrobovT1, KadigrobovT2}. In order to obtain these transitions, one should take into account that the electron momentum should be replaced with $\left(\hat{\bm{p}} - \frac{e}{c} \bm{A}_\omega\right)$ in $J$. This leads to additional operator of interaction of electron and electromagnetic wave linear in $\bm{E}_\omega$
\begin{equation} \label{Eq1_em2}
\hat{H}_{e-m}' = -i \frac{e}{\omega} \left(\bm{E}_\omega \cdot \frac{\partial J}{\partial \bm{p}}\right) \left(\hat{\bm{\sigma}} \cdot \bm{M}\right).
\end{equation}
The interaction operator (\ref{Eq1_em2}) flips spins and therefore leads to electron transitions. This mechanism of electromagnetic wave emission was studied in some later papers \cite{Ustinov, ZilbermanT}. Experiments on electromagnetic wave generation in non-uniform ferromagnets were performed inspired by these investigatios \cite{KadigrobovE, ZilbermanE1, ZilbermanE2}. Additional modifications such as anisotropic exchange interaction were considered in order to improve the wave generation properties \cite{???}. However it is hard to separate the mentioned mechanism from other possible in the existing experiments. Another important property of this mechanism is its spin-orbit nature. Indeed, only exchange interaction is taken into account in the hamiltonian (\ref{Eq1_ham}). But the dependence of $J$ on electron momentum may rise only from spin-orbit interaction in the subsystem of localized d- or f-electrons. Moreover, this mechanism exists only in non-collinear magnetic systems.

There are mechanisms of electron spin-flip transitions in ferromagnets different from one described in the previous paragraph. The most simple mechanism is caused by the magnetic field of the wave. The magnetic field provides transitions through the Zeeman term \cite{Landau3}
\begin{equation} \label{Eq1_Z}
\hat{H}_Z = \mu_0 g \left(\hat{\bm{\sigma}} \cdot \bm{B}_\omega\right)
\end{equation}
in hamiltonian where $\bm{B}_\omega$ is the magnetic field of the wave, $g$ is the electron g-factor that is supposed to be equal to $2$, $\mu_0$ is the Bohr magneton. Another mechanism that exists in  homogeneously magnetized ferromagnet is caused by the spin-orbit interaction. In this paper the Rashba coupling \cite{Tatara, Rashba} is considered
\begin{equation} \label{Eq1_R}
\hat{H}_R = i \left(\bm{\alpha}_R \cdot \left[\nabla \times \hat{\bm{\sigma}}\right]\right).
\end{equation}
This interaction usually appears at the surfaces, and the Rashba vector $\bm{\alpha}_R$ is parallel to the surface normal. Finally, a fully exchange mechanism exists in non-collinearly magnetized ferromagnets due to the connection between spin and spatial coordinates in such ferromagnets. This mechanism was theoretically demonstrated recently \cite{KarashtinJETPL} and is studied insufficiently. We consider the magnetic helicoid (Bloch type spiral):
\begin{equation} \label{Eq1_Bloch}
\bm{M} = \bm{e}_x cos qz + \bm{e}_y sin qz,
\end{equation}
where $q$ is inversely proportional to the spiral step, and the Cartesian coordinate system with the z-axis along the spiral axis is chosen. This type of magnetization structure is realized in holmium at low temperature \cite{Koehler}. In this material, electromagnetic wave absorption with spin flip electron transitions is known \cite{Weber, KarashtinJETP}. This phenomenon is inverse to the electromagnetic wave generation which is the subject of current paper.

In order to obtain electromagnetic wave generation, one has to inject non-equilibrium spin into the upper subband. There are several ways to do this \cite{KarashtinFTT}. The most suitable for ferromagnetic metals are based on spin pumping effect \cite{Tserkovnyak, Sinova, Fert, Mizukami, Wang} and on the injection of spin-polarized current \cite{Slonczewski, Berger, Zutic}. The spin pumpig effect is usually quite weak. Besides, it is hard to control the magnitude of spin current. On the other hand, injection of non-equilibrium spin by a spin-polarized electric current is realized in a simple system consisting of two ferromagnets divided by a nonmagnetic interlayer. The voltage is applied to this system which causes the electric current to flow from one ferromagnet (spin source) to the other (active region). The amount of injected spin may be varied widely by changing the applied voltage. Therefore we consider this mechanism of non-equilibrium spin injection.

The paper is organized as follows. First, we perform calculations of electron transition probabilities for four mentioned mechanisms. The transition probabilities are estimated for realistic parameters and are compared to each other in Section~\ref{Sec2}. After that, we consider a simple model in which non-equilibrium spin is injected into the ferromagnet by a spin-polarized electric current through a tunnel barrier in Section~\ref{Sec3}. The estimations of the radiation power and threshold electric current are performed.

\section{\label{Sec2}Electron transition probabilities}
We describe the conduction electrons by the hamiltonian (\ref{Eq1_ham}) which takes into account the exchange coupling in the s-d model approach. The magnetization $\bm{M}$ may either be constant or depend on coordinates (see (\ref{Eq1_Bloch})). Different types of interactions additionally taken into account lead to different probabilities of electron transitions between spin subbands. While the Rashba and Zeeman coupling may be taken into account in perturbation theory, non-collinear magnetic system has sufficiently different spin subbands.

Taking uniform $\bm{M} = \bm{e}_z$, it is easy to find wave functions in the form (\ref{Eq1_wf}) and corresponding energy of electrons:
\begin{equation} \label{Eq2_energy}
\varepsilon_{\pm} = \frac{\hbar^2 \bm{k}^2}{2 m_e} \pm J.
\end{equation}
In order to calculate the probability of electron transitions between spin subbands, we use the Fermi golden rule \cite{Landau3}
\begin{equation} \label{Eq2_Fermi}
W^\pm_{\bm{kk}'} = \frac{2 \pi}{\hbar} \left|H^\pm_{\bm{kk}'}\right|^2 \frac{\Delta / \pi}{\left(\Delta \varepsilon - \hbar \omega\right)^2 + \Delta^2}
\end{equation}
where spin flip processes due to reasons other than interaction with electromagnetic wave (e.g. scattering on magnetic impurities) are taken into account, $\tau_s = 2 \pi \hbar / \Delta$ being the spin relaxation time; $H^\pm_{\bm{kk}'} = \left<\psi_+\left(\bm{k}'\right) | \hat{H}^{int} | \psi_-\left(\bm{k}\right)\right>$ is the matrix element of the hamiltonian of interaction of electrons with electromagnetic wave $\hat{H}^{int}$; $\left(\bm{k}, -\right)$ and $\left(\bm{k}', +\right)$ are the initial and final states; $\Delta \varepsilon = \varepsilon_+\left(\bm{k}'\right) - \varepsilon_-\left(\bm{k}\right)$ is the energy difference between these states.

Electron transitions due to magnetic field of the wave are described by the interaction hamiltonian (\ref{Eq1_Z}). The magnetization is supposed to be uniform. Using (\ref{Eq2_Fermi}) we arrive at
\begin{equation} \label{Eq2_WB}
W^\pm_B = \frac{2}{\hbar} \left(\frac{e \hbar \left|\left[\bm{B} \times \bm{M}\right]\right|}{m_e c}\right)^2 \frac{\Delta}{\left(\Delta \varepsilon - \hbar \omega\right)^2 + \Delta^2} \delta\left(\bm{k} - \bm{k}'\right),
\end{equation}
$\delta\left(\right)$ is the Dirac delta-function.
It is seen that only the magnetic field orthogonal to the magnetization $\bm{M}$ participates in intersubband spin transitions. The quasimomentum (or electron wave vector) is conserved here, just as in all mechanisms  of electron transitions considered below.

\begin{figure}[t]
\center{\includegraphics[width=0.8 \textwidth, keepaspectratio=true]{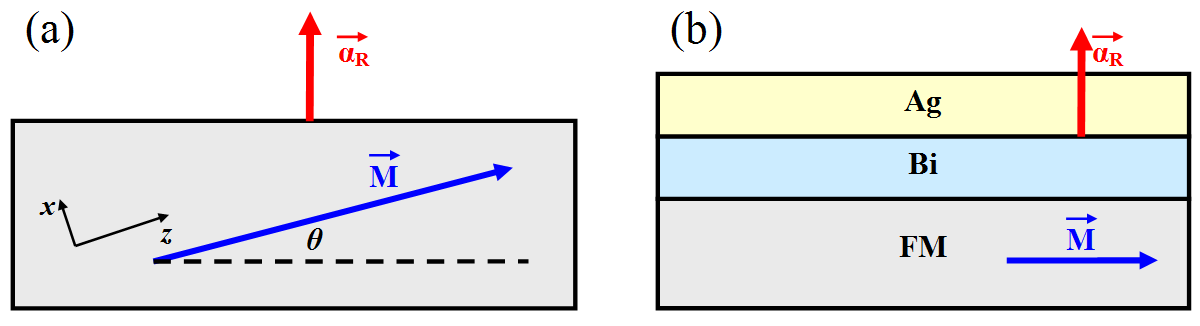}}
\caption{\label{Fig1} (Color online) (a) Uniformly magnetized system with Rashba coupling. (b) A possible way to increase the Rashba coupling at the interface of a ferromagnet.}
\end{figure}

In order to find probability of electron transitions due to Rashba spin-orbit coupling, it is necessary to take into account both (\ref{Eq1_R}) and (\ref{Eq1_em}). The coordinate system is chosen so that $\bm{M} = \bm{e}_z$, and $\bm{\alpha}_R = \alpha_R \left(\bm{e}_x cos \theta + \bm{e}_z sin \theta \right)$ (see Figure~\ref{Fig1}a) thus describing general case.
We restrict ourselves to linear in $\alpha_R$ order. In this approximation the energy and wave functions are
\begin{equation}\label{Eq2_energyR}
\varepsilon_\pm = \frac{\hbar^2 \bm{k}^2}{2 m_e} \pm \left(J - \alpha_R k_y cos \theta\right),
\end{equation}
\begin{equation} \label{Eq2_wfRp}
\psi_+ = exp\left(i \bm{k r}\right)
\begin{pmatrix}
1  \\
-\frac{\alpha_R \left(\left(k_y - i k_x\right) sin\theta - i k_z cos \theta \right)}{2J}
\end{pmatrix},
\end{equation}
\begin{equation} \label{Eq2_wfRm}
\psi_- = exp\left(i \bm{k r}\right)
\begin{pmatrix}
\frac{\alpha_R \left(\left(k_y + i k_x\right) sin\theta + i k_z cos \theta \right)}{2J}  \\
1
\end{pmatrix}.
\end{equation}
Substituting these wave functions into (\ref{Eq2_Fermi}) with (\ref{Eq1_em}) as the interaction hamiltonin we see that the momentum is conserved and the matrix element of transitions is zero. Note that this property remains in higher orders in $\alpha_R$. Thus there are no spin-flip electron transitions. However the Rashba hamiltonian (\ref{Eq1_R}) itself depends on the electron momentum. This momentum $\hat{\bm{p}} = -i \hbar \nabla$ should be changed to $\hat{\bm{p}} - \frac{e}{c} \bm{A}_\omega$. In turn, this leads to another interaction hamiltonian
\begin{equation} \label{Eq2_em3}
\hat{H}_{e-m}'' = \frac{i e}{2 \hbar \omega} \left(\bm{\alpha_R} \cdot \left[\hat{\bm{\sigma}} \times \bm{E}_\omega \right] \right).
\end{equation}
The hamiltonian (\ref{Eq2_em3}) is linear in Rashba interaction itself. Therefore its matrix elements should be found with the use of wave functions (\ref{Eq1_wf}). After substituting them into (\ref{Eq2_Fermi}) we have
\begin{equation} \label{Eq2_WR}
W^\pm_R = \frac{2}{\hbar} \left(\frac{e \alpha_R}{2 \hbar \omega}\right)^2 \left(E_\tau^2 + E_y^2 sin^2 \theta\right) \frac{\Delta}{\left(\Delta \varepsilon - \hbar \omega\right)^2 + \Delta^2} \delta\left(\bm{k} - \bm{k}'\right),
\end{equation}
where $E_\tau$ is the component of wave electric field which is orthogonal to $\bm{\alpha}_R$. It is obvious that in such approach the electron transition probability is proportional to the square of the Rashba constant in the lowest order. The electrons transitions go most effectively when $sin^2 \theta = 1$, i.e. the Rashba vector $\bm{\alpha}_R$ is either parallel or antiparallel to magnetization $\bm{M}$. However it is of the same order for $\bm{\alpha}_R$ perpendicular to $\bm{M}$.

We consider two other mechanisms that exist due to exchange interaction in non-collinear magnetic state. The mechanism that follows from dependence of the exchange constant $J$ on electron momentum is described by the interaction hamiltonian ($\ref{Eq1_em2}$). It is important that if the magnetization is uniform ($\bm{M} = \bm{e}_z$) this hamiltonian contains only $\hat{\sigma}_z$ and therefore it does not flip spins. In literature this mechanism was suggested and studied for a non-collinear system that consists of two layers uniformly magnetized in different directions \cite{KadigrobovT1, ZilbermanT}. We study this mechanism for the helicoidal magnetization distribution (\ref{Eq1_Bloch}).

Exact solutions of the Schrodinger equation with hamiltonian (\ref{Eq1_ham}) is known from literature \cite{Calvo, Nagaev}:
\begin{equation} \label{Eq2_wfSp}
\psi_+ = \frac{1}{\sqrt{1 + v^2}} e^{-i \frac{\varepsilon_+}{\hbar} t + i \bm{k} \bm{r}} 
\begin{pmatrix}
v e^{-i \frac{q}{2} z} \\
e^{i \frac{q}{2} z}
\end{pmatrix},
\end{equation}
\begin{equation} \label{Eq2_wfSm}
\psi_- = \frac{1}{\sqrt{1 + v^2}} e^{-i \frac{\varepsilon_-}{\hbar} t + i \bm{k} \bm{r}} 
\begin{pmatrix}
e^{-i \frac{q}{2} z} \\
- v e^{i \frac{q}{2} z}
\end{pmatrix},
\end{equation}
where $q$ is the wave vector that is determined by the spiral step, as before, and the constant $v\left(k_z\right)$ is determined by
\begin{equation} \label{Eq2_delta}
v = \frac{j}{k_z q + \sqrt{j^2 + k_z^2 q^2}} \equiv \frac{-k_z q + \sqrt{j^2 + k_z^2 q^2}}{j}
\end{equation}
where we introduce the notation $j = \frac{2 m_e}{\hbar^2} J$. The eigenstates (\ref{Eq2_wfSp}), (\ref{Eq2_wfSm}) correspond to energy defined as
\begin{equation} \label{Eq2_Sen}
\varepsilon_\pm = \frac{\hbar^2}{2 m_e} \left(\bm{k}^2 + \frac{q^2}{4} \pm \sqrt{j^2 + k_z^2 q^2}\right).
\end{equation}
After calculating the matrix elements of the hamiltonian (\ref{Eq1_em2}) on the wave functions (\ref{Eq2_wfSp}), (\ref{Eq2_wfSm}) we get the electron transitions probability
\begin{equation} \label{Eq2_WJ}
W^\pm_J = \frac{2}{\hbar} \left( \frac{e \hbar}{m_e \omega} \left(\bm{E} \cdot \frac{\hbar k_z}{J} \frac{\partial J}{\partial \bm{p}}\right) \right)^2 q^2 \frac{J^2}{\Delta \varepsilon^2}
\frac{\Delta}{\left(\Delta \varepsilon - \hbar \omega\right)^2 + \Delta^2} \delta\left(\bm{k} - \bm{k}'\right).
\end{equation}
This probability is proportional to $\left(q l_\omega\right)^2$ where $l_\omega \propto E$ is the magnitude of classic oscillations of electrons in wave electric field in the direction in which $J$ depends on $\bm{k}$. Importantly, smaller scale of magnetization change leads to bigger effect. This proves the fact that this effect of electrons transitions follows from non-collinear magnetic structure. It is also important to note that the electron energy change $\Delta \varepsilon$ depends on $k_z$ for magnetic helicoid. Therefore the spin relaxation processes that lead to energy uncertainty are important here.

A mechanism that is fully exchange exists in non-collinear ferromagnet due to the minimal coupling described by (\ref{Eq1_em}). It does not need the dependence of the exchange constant $J$ on the electron quasimomentum. The electron transitions probability determined by (\ref{Eq2_Fermi}) for the interaction hamiltonian (\ref{Eq1_em}) with the wave functions (\ref{Eq2_wfSp}), (\ref{Eq2_wfSm}) is
\begin{equation} \label{Eq2_Whel}
W^\pm_{hel} = \frac{2}{\hbar} \left(\frac{e \hbar E_z}{2 m_e \omega}\right)^2 q^2 \frac{J^2}{\Delta \varepsilon^2} \frac{\Delta}{\left(\Delta \varepsilon - \hbar \omega\right)^2 + \Delta^2} \delta\left(\bm{k} - \bm{k}'\right).
\end{equation}
This probability is also proportional to $\left(q l_\omega\right)^2$, but the magnitude of classic oscillations along the direction of magnetization change are important here. Thus, (\ref{Eq2_WJ}) and (\ref{Eq2_Whel}) lead to different polarization properties.

The absolute value of probability is proportional to wave intensity and depends on polarization. Therefore in the following part of this Section we compare all four obtained electron transition probabilities (\ref{Eq2_WB}), (\ref{Eq2_WR}), (\ref{Eq2_WJ}), and (\ref{Eq2_Whel}) to each other. The probability $W^\pm_{hel}$ that exists in magnetic spiral is taken as a reference since it is fully exchange. If we take $\hbar \omega \approx 2 J$ thus supposing the resonant character of electron transitions the ratio $W^\pm_B / W^\pm_{hel}$ may be roughly determined as $W^\pm_B / W^\pm_{hel} \approx \left(\frac{4 k_\omega}{q}\right)^2$, where $k_\omega = \frac{\omega}{c}$ is the wave vector of the electromagnetic wave in vacuum. Taking the parameters of holmium \cite{Koehler} $q \approx 10^7 cm^{-1}$ (which corresponds to the spiral step $3.5nm$), $J \approx 0.185 eV$ we obtain the estimation $W^\pm_B / W^\pm_{hel} \approx 5.6 \cdot 10^{-5}$. For another example of helical ferromagnet, manganese silicene, $q$ is approximately 6 times smaller, and therefore $W^\pm_B / W^\pm_{hel} \approx 1.8 \cdot 10^{-3}$. Thus, the magnetodipole electron transitions due to Zeeman term have very low probability and usually may be neglected. The transition probabilities depend on wave frequency in different ways, but the resonance should be shifted to very high frequency in order to obtain $W_B^\pm$ comparable to other transition probabilities. This does not correspond to any real system parameters. We therefore omit transitions due to the magnetic field of the wave below.

Comparing the Rashba mechanism to the exchange mechanism in helicoid, again for $\hbar \omega \approx 2 J$, gives $W^\pm_R / W^\pm_{hel} \approx \left(\frac{m_e \alpha_R}{\hbar^2 q}\right)^2$. The Rashba interaction usually appears at interfaces. The value of $\alpha_R$ for ferromagnets may be estimated from known value of the Dzyaloshinskii-Moriya interaction e.g. in a [Co/Pt] multilayer \cite{Udalov}. This estimation gives $\alpha_R \approx 1 peV \cdot m$. Then $W^\pm_R / W^\pm_{hel} \approx 0.013$ for holmium and $\approx 0.4$ for MnSi. Therefore the Rashba interaction mechanism should be important even for simple ferromagnet interfaces. If we consider a three-layer system in which an interface on two non-magnetic materials with strong Rashba coupling is placed close to the ferromagnet (see Figure~\ref{Fig1}b), the Rashba coupling may become the main mechanism of electron transitions between spin subbands. For example, $\alpha_R \approx 305 peV \cdot m$ at a Bi / Ag interface \cite{Fert}. So in a  ferromagnet / Bi / Ag system with a thin Bi layer $W^\pm_R / W^\pm_{hel} \approx 4$ even if holmium is taken to calculate $W^\pm_{hel}$ in which the spiral period is extremely small.

Two mechanisms of interband electron transitions in non-collinear ferromagnets defined by (\ref{Eq2_WJ}) and (\ref{Eq2_Whel}) have different symmetry with respect to the wave polarization: the first one is determined by the dependence of $J$ on $\bm{k}$ and therefore by the crystallographic structure of the ferromagnet, while the second one is determined by the direction of magnetization change. The direction of change of magnetic moment in natural non-collinear ferromagnets such as holmium is strictly connected to their crystallographic directions and therefore these two mechanisms are similar in such natural ferromagnets. However in artificial ferromagnets the situation may change. In principle, in a polycrystalline magnetic structure with random directions of crystallographic axes the mechanism defined by (\ref{Eq2_WJ}) may vanish while (\ref{Eq2_Whel}) depends only on magnetic structure and is therefore may exist. In order to compare these two mechanisms quantitatively, we use the estimation from literature \cite{KadigrobovT1, KadigrobovT2}
\begin{equation} \label{Eq2_estim}
\left|\frac{\partial J}{\partial \bm{p}}\right| \approx \frac{J}{p_0},
\end{equation}
where $p_0 = \frac{\hbar}{a}$, and $a \approx 10^{-8} cm$ is the lattice constant. Then the ratio $W^\pm_J / W^\pm_{hel} \approx \left(k_z a\right)^2$. For $k_z \approx k_f$ and the Fermi energy $\varepsilon_f = \frac{\hbar^2 k_f^2}{2 m_e} \approx 5eV$ we have $W^\pm_J / W^\pm_{hel} \approx 1$. However the estimation (\ref{Eq2_estim}) is very optimistic, and besides $1$ is obtained for maximal value of $k_z^2$. Therefore it seems that the electron transitions in a non-collinear ferromagnet that exist for constant $J$ would be more effective than that which appear from the dependence of $J$ on the electron quasimomentum. At the same time, both mechanisms are of the same order of value and therefore should be taken into account simultaneously.

The electron transition probabilities generally depend on the initial and final electron quasimomentum. In order to quantitatively determine the effectiveness of electron transitions we need to average them over electron states. Therefore we consider a simple model of non-equilibrium spin injection into the ferromagnet in the following section. The power of electromagnetic wave radiation is also estimated.

\section{\label{Sec3}Emission rate and electromagnetic wave radiation power}
The electromagnetic wave radiation power may be found knowing the stimulated emission rate $R_{st}$ which is defined as
\begin{equation} \label{Eq3_Rst}
R_{st} = \int{d\bm{k} d\bm{k}' W^\pm_{\bm{kk}'} \left(f_+\left(\bm{k}'\right) - f_-\left(\bm{k}\right)\right)},
\end{equation}
where $f_\pm$ are the electron distribution functions in two subbands. This emission rate is an average characteristic of electron transitions. It is proportional to the intensity of electromagnetic wave, or the density of photons $N_p$:
\begin{equation} \label{Eq3_G}
R_{st} = G N_p,
\end{equation}
where $G$ depends on the electron distribution functions in two spin subbands. These distribution functions are non-equilibrium and are determined by the model of non-equulibrium spin injection into the system.  Therefore we first consider a simple model of injection of spin into the ferromagnet and then study the emission properties in this system.

\subsection{Spin injection into a ferromagnet}
\begin{figure}[t]
\center{\includegraphics[width=0.8 \textwidth, keepaspectratio=true]{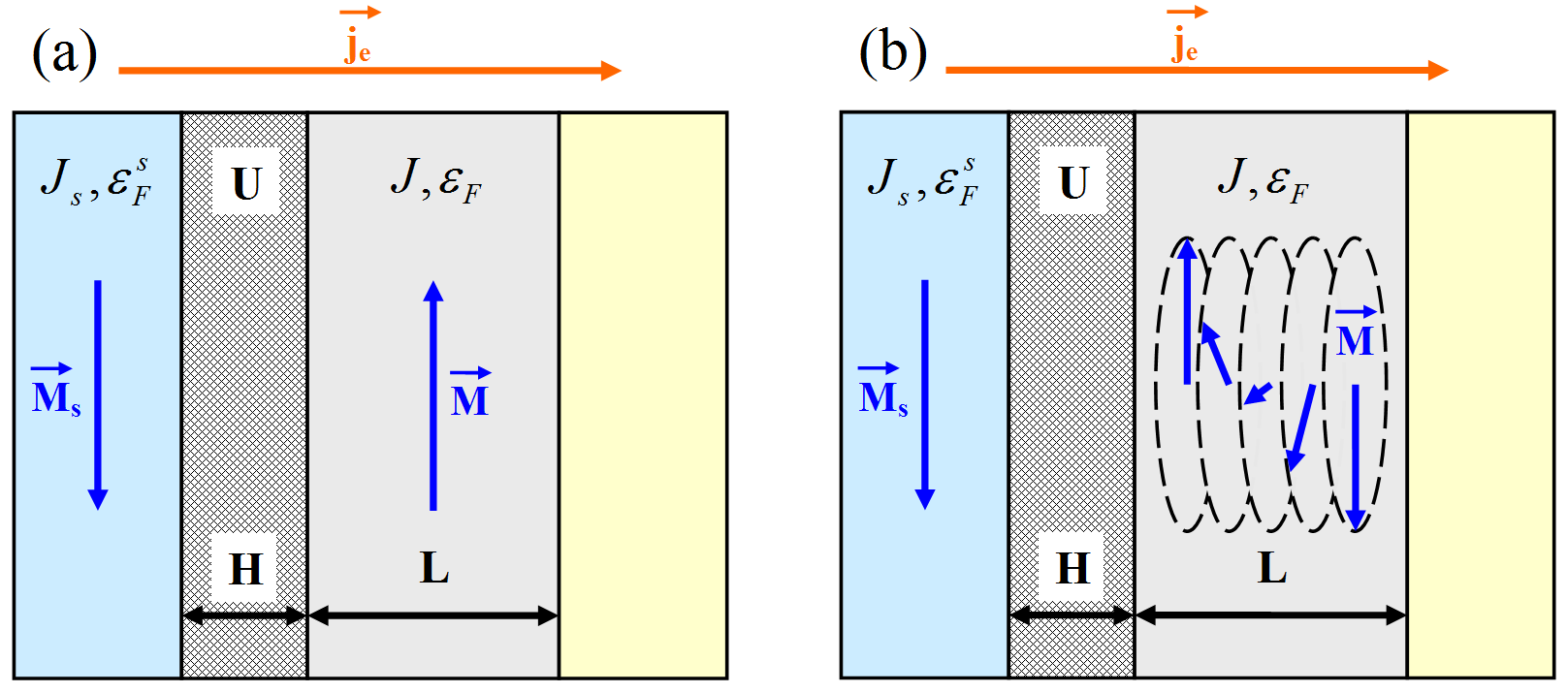}}
\caption{\label{Fig2} (Color online) Spin injection into a ferromagnet (active region with $\bm{M}, J, \varepsilon_F$) from another ferromagnet (spin source with $\bm{M_s}, J_s, \varepsilon_F^s$) for (a) uniform and (b) helicoidal active region.}
\end{figure}

We suppose that the electrons are injected into the ferromagnet (active region) from another ferromagnet by electric current through a tunnel barrier with constant height $U$ and thickness $H$ (see Figure~\ref{Fig2}).
The Fermi level in the ferromagnet that acts as a spin source $\varepsilon_F^s$ is supposed to be greater than that in another ferromagnet $\varepsilon_F$ which is the active region. In equilibrium, electrons tunnel from the spin source into the active region which leads to electric polarization of boundaries alike the contact potential difference. These tunneling processes are not important in the framework of current paper and therefore are not considered. We suppose that the voltage $V$ is applied to the system. This potential difference drops on the the tunnel barrier. The probability of electron tunnelling through the barrier linear in voltage $V$ is determined by the equation
\begin{equation} \label{Eq3_P}
P^{tunn}_{\pm} =A\left(U\right) \left(\varepsilon - \varepsilon_\perp \pm J_s\right),
\end{equation}
where $J_s$ is the exchange constant in the spin source, $A\left(U\right)$ is determined as
\begin{equation} \label{Eq3_Adef}
A\left(U\right) = 2 H \sqrt{\frac{2 m_e}{\hbar^2} U} \frac{e V}{U^2} \exp{\left(-2 H \sqrt{\frac{2 m_e}{\hbar^2} U}\right)},
\end{equation}
and we suppose that the barrier height $U$ is much greater than the Fermi energy $\varepsilon_F^s$ and the energy $e V H$ gained by the electron in the electric field created by the voltage. The quantum number $k$ in the direction along the normal to boundaries is mixed with the electron spin . Therefore we use quantum numbers $\varepsilon$ which is the electron energy and $\varepsilon_\perp = \frac{\hbar^2 k_\perp^2}{2 m_e}$ where $k_\perp$ is the quasimomentum along the surface of the boundaries (the direction of this quasimomentum is not important and therefore a simple integration over it leads to multiplication by $2 \pi$).

We suppose that the spin quantization axis is chosen as in the active region. The spin source magnetization has the direction opposite to the magnetization in the active region (see Figure \ref{Fig2}a), so that more electrons are injected into upper spin subband than into lower one. For the helicoidal magnetization the system is chosen so that the axis of the helicoid is perpendicular to the tunnel barrier surface. The magnetization of the spin source is directed oppositely to the magnetization of the active region at the boundary (Figure~\ref{Fig2}b). In this case the electrons injected with the spin parallel or antiparallel to the magnetization of spin source may be roughly considered as injected to minority or majority spin subband correspondingly which may lead to population inversion \cite{KarashtinFTT}.
The probability (\ref{Eq3_P}) depends on the spin of the electron. It is non-zero for the energy of electron $\varepsilon_F < \varepsilon < \varepsilon_F^s$ because all states with $\varepsilon < \varepsilon_F$ are occupied by other electrons in the active region and therefore tunneling to these states is impossible. On the other hand, $\varepsilon<\varepsilon_F^s$ is governed by the fact that there are no electrons with energy greater than $\varepsilon_F^s$ in the spin source. The range for $\varepsilon_\perp$ is determined by the demand of real $k$ along the normal to boundary surface. For the lower subband (``-'') $0<\varepsilon_\perp<\varepsilon - J_s$ which is determined by the spin source. For the upper subband (``+'') $0<\varepsilon_\perp<\varepsilon - J$ which is determined by the demand of real $k$ in the active region. Note that there are electrons with $\varepsilon - J < \varepsilon_\perp < \varepsilon + J_s$ in the spin source but they are reflected from the boundary.

The spin-polarized electric current is determined by the electron momentum averaged with the tunneling probability (\ref{Eq3_P}). After a simple calculation we arrive at
\begin{equation} \label{Eq3_js}
j_{e \pm} = A\left(U\right) \frac{\pi e m_e}{3 \hbar^3} \left(\left(\varepsilon_F^s\right)^3 - \varepsilon_F^3\right) \left( 1 \pm \frac{3}{2} \frac{\varepsilon_F^s + \varepsilon_F}{\left(\varepsilon_F^s\right)^2+\varepsilon_F^2+\varepsilon_F \varepsilon_F^s} \left(J_s - J\right) \right).
\end{equation}
Taking into account that $j_{e+} + j_{e-} = j_e$ we find the constant $A\left(U\right)$ as
\begin{equation} \label{Eq3_A}
A\left(U\right) = j_e \frac{3 \hbar^3}{2 \pi e m_e \left( \left(\varepsilon_F^s\right)^3 - \varepsilon_F^3 \right)}.
\end{equation}
Equation (\ref{Eq3_A}) together with (\ref{Eq3_P}) determines the tunneling probability via the electric current density $j_e$.

The electron density in two spin subbands in the active region is denoted as $N_\pm$ and the correction to this density due to spin polarized current is $\delta N_\pm$. We suppose that the total electron number does not change in time in the active region, i.e. the electrons are injected via the tunnel barrier and leave on the other side of the active region. In this case $\delta N_\pm = \delta N_0 \pm \delta N$, where the total number of injected electrons $2 \delta N_0$ is constant and $\delta N$ determines the electron spin polarization. Besides, we describe the system by the electron density averaged over the thickness of the active region. However we take into account spin relaxation inside the active region. If the spin relaxation length is $\lambda_s$ and the part of electrons injected into upper (lower) spin subband with respect to total number of injected electrons is $a_\pm = \frac{j_{e \pm}}{j_e}$ then the part of the electrons that leave the active region with the same spin is $\left(a_\pm - \frac{1}{2}\right)\exp{\left(-\frac{L}{\lambda_s}\right)} + \frac{1}{2}$ where $L$ is the thickness of active region. We then have for the electron density averaged over the active region thickness
\begin{equation} \label{Eq3_dN}
\dot{\delta N_+} = -\dot{\delta N_-} = \dot{\delta N} = \frac{j_e}{e L} \eta \left(1 - \exp{\left(-\frac{L}{\lambda_s}\right)}\right) - \frac{\delta N}{\tau_s},
\end{equation}
where $\tau_s$ is the spin relaxation time as before, and the efficiency of spin injection $\eta$ is determined as
\begin{equation} \label{Eq3_eta}
\eta = \frac{1}{2} \frac{j_+ - j_-}{j_+ + j_-} = \frac{3}{4} \frac{\varepsilon_F^s + \varepsilon_F}{\left(\varepsilon_F^s\right)^2 + \varepsilon_F^2 + \varepsilon_F^s \varepsilon_F} \left(J_s - J\right).
\end{equation}
This corresponds to previously used equation for spin-polarized electron density \cite{KadigrobovT1, KadigrobovT2, KarashtinJETPL} where $\eta$ was introduced phenomenologically. It should be noted that the spin injection efficiency is determined by both the exchange constant of spin source $J_s$ and the exchange constant of active region $J$. If $J > J_s$ then there is no population inversion, i.e. more electrons are injected into lower spin subband than into upper one. In order to have $\eta >0$ one needs to inject spins from a stronger ferromagnet than that is used as the active region, for which the inequality $J_s > J$ is satisfied. The equation (\ref{Eq3_dN}) determines the stationary non-equilibrium state with no stimulated emission:
\begin{equation} \label{Eq3_eq}
\delta N^* =  \frac{j_e \tau_s}{e L} \eta \left(1 - \exp{\left(-\frac{L}{\lambda_s}\right)}\right).
\end{equation}
In this state the injected non-equilibrium spin is compensated by spin relaxation in the active region.

In the existing literature \cite{KadigrobovT1, KadigrobovT2, ZilbermanT, KarashtinJETPL} it is implied that all injected electrons immediately relax to the lowest vacant states inside the spin subband, thus supposing that the energy relaxation time $\tau_e << \tau_s$ and besides $\tau_e << \frac{L}{v_F}$. However typical time of electron pass through the active region $\frac{L}{v_F}$ is much smaller than $\tau_e$ for the thickness of the active region $L \sim 10nm$ (the electromagnetic wave emission is accompanied by the change of electron energy but these processes are supposed to give a small correction to $\tau_e$) \cite{KarashtinJPCM}. Therefore in current paper we suppose that electron energy does not relax inside the active region (i.e. the electron transitions with photon emission are supposed to give a small correction to the distribution function which is beyond the scope of our consideration). The non-equilibrium correction to electron distribution function is determined by the probability of electron injection (\ref{Eq3_P}):
\begin{equation} \label{Eq3_distrp}
\delta f_+ =  j_e \frac{3 \hbar^3 \left(\varepsilon - \varepsilon_\perp + J_s\right)}{4 \pi e m_e \left( \left(\varepsilon_F^s\right)^3 - \varepsilon_F^3 \right)} \frac{\lambda_s}{L}\left(1 - e^{-\frac{L}{\lambda_s}}\right), \varepsilon \in \left(\varepsilon_F, \varepsilon_F^s\right), \varepsilon_\perp \in \left(0, \varepsilon-J\right),
\end{equation}
\begin{equation} \label{Eq3_distrm}
\delta f_- =  j_e \frac{3 \hbar^3 \left(\varepsilon - \varepsilon_\perp - J_s\right)}{4 \pi e m_e \left( \left(\varepsilon_F^s\right)^3 - \varepsilon_F^3 \right)} \frac{\lambda_s}{L}\left(1 - e^{-\frac{L}{\lambda_s}}\right), \varepsilon \in \left(\varepsilon_F, \varepsilon_F^s\right), \varepsilon_\perp \in \left(0, \varepsilon-J_s\right),
\end{equation}
where we take into account the spin relaxation processes and average the distribution function over the thickness of the active region and also take into account that only electrons with positive $k$ from spin source to active region participate in $\delta f_\pm$. Note that the equilibrium electron distribution function is determined as
\begin{equation} \label{Eq3_f0}
f_{0 \pm} = 2,  \varepsilon \in \left(\pm J, \varepsilon_F\right), \varepsilon_\perp \in \left(\pm , \varepsilon \mp J\right),
\end{equation}
where we take into account that there are electrons with two signs of $k_z$ for each energy level, and the nonequilibrium distribution function is $f_\pm = f_{0 \pm} + \delta f_\pm$. It is important to note that the emission rate (\ref{Eq3_Rst}) contains the difference of electron densities at different energy rates $f_+\left(\varepsilon + \Delta \varepsilon\right)-f_-\left(\varepsilon\right)$, where $\Delta \varepsilon$ is the energy difference between two spin subbands.

We see from (\ref{Eq3_distrp}), (\ref{Eq3_distrm}) that the model of spin injection into the active region is very important because it determines the non-equilibrium electron distribution function. In our model, the result is obtained for the non-equilibrium distribution function in terms of $j_e$ and thus does not directly contain the barrier height $U$.

\subsection{Emission rate and critical current}
Knowing the electron distribution function (\ref{Eq3_distrp}), (\ref{Eq3_distrm}) and the probability of electron transition between two spin subbands we may find the stimulated emission rate with the use of (\ref{Eq3_Rst}). There are two contributions into $R_{st}$. The first one is due to equilibrium part of distribution function. Electrons with $\varepsilon \in \left(\varepsilon_F - 2J, \varepsilon_F\right)$ may transfer from the lower spin subband into the upper one because states with the same momentum have $\varepsilon > \varepsilon_F$ in the upper subband and therefore are free. Obviously, this gives negative contribution to $R_{st}$ which corresponds to electromagnetic wave absorption.

The energy difference between two spin subbands $\Delta \varepsilon$ is constant and equal to $2J$ for uniform magnetization $\bm{M}$. For the magnetic helicoid $\Delta \varepsilon = \frac{\hbar^2}{m_e} \sqrt{j^2 + k_z^2 q^2}$ and thus depends on $k_z$. The dependence of $\Delta \varepsilon$ should be taken into account for $k_z \sim k_F$. For holmium, $q \sim 1.8 \cdot 10^7 cm^{-1}$ and $J \approx 0.185eV$ \cite{Weber, Koehler, KarashtinJETP} which gives $j \sim 0.48 \cdot 10^{15} cm^{-2}$, and taking $\varepsilon_F \approx 1eV$ we have $k_F q \sim 0.95 \cdot 10^{15} cm^{-2}$. However the approximation $\Delta \varepsilon \approx 2 J$ is sopmetimes useful because it allows to integrate everything exactly. Besides, it is correct for the mechanism that appears due to Rashba coupling in a uniform ferromagnet. Therefore we use this approximation at first and then discuss the result and make corrections.

Restricting ourselves to the lowest nonzero order in $\frac{J}{\varepsilon_F}$ and $\frac{J_s}{\varepsilon_F}$, we obtain
\begin{equation} \label{Eq3_R1R}
R_{st\,R}^{\left(1\right)} = - \left(\frac{e \left[\bm{\alpha}_R \times \bm{E}\right]}{\hbar \omega}\right)^2 \frac{2 \pi \left(2 m_e\right)^{3/2}}{\hbar^4} \frac{\Delta}{\left(\hbar \omega - 2 J\right)^2 + \Delta^2} \frac{J}{\varepsilon_F} \varepsilon_F^{3/2},
\end{equation}
\begin{equation} \label{Eq3_R1hel}
R_{st\,hel}^{\left(1\right)} = - \left(\frac{e E_z}{\hbar \omega}\right)^2 \frac{2 \pi q^2}{\sqrt{2 m_e}} \frac{\Delta}{\left(\hbar \omega - 2 J\right)^2 + \Delta^2} \frac{J}{\varepsilon_F} \varepsilon_F^{3/2},
\end{equation}
for all $W_{\bm{kk}'}^\pm$ except for $W_J^\pm$ (for $W_R^\pm$, we suppose that the Rashba vector is perpendicular to $\bm{M}$ as shown in Figure~\ref{Fig2}a). For the mechanism determined by $W_J^\pm$ the transition probability is proportional to $k_z^2$ and therefore we obtain a different result
\begin{equation} \label{Eq3_R1J}
R_{st\,J}^{\left(1\right)} = -\left(\frac{e \left(\bm{E} \cdot \frac{\hbar k_F}{J} \frac{\partial J}{\partial \bm{p}}\right)}{\hbar \omega}\right)^2 \frac{8 \pi q^2}{3 \sqrt{2 m_e}} \frac{\Delta}{\left(\hbar \omega - 2 J\right)^2 + \Delta^2} \frac{J}{\varepsilon_F} \varepsilon_F^{3/2},
\end{equation}
The part of $R_{st}$ determined by (\ref{Eq3_R1R})--(\ref{Eq3_R1J}) depends only on the parameters of the active region. It does not depend on the electric current $j_e$. This is obvious since it corresponds to absorption of light by equilibrium electrons. In previous papers this part was not taken into account directly. However we show below that it is very important.

The contribution into $R_{st}$ which appears due to non-equilibrium spin injection ($R_{st}^{\left(2\right)}$) is determined by $\left(\delta f_+\left(\varepsilon + \Delta \varepsilon\right) - \delta f_-\left(\varepsilon\right)\right)$. It consists of two terms: one appears due to electron transitions from the upper subband into the lower one and is positive; the other appears due to backward electron transitions close to $\varepsilon_F^s$ where $\delta f_+\left(\varepsilon + \Delta \varepsilon\right) = 0$ and is negative. After a simple calculation we arrive at
\begin{equation} \label{Eq3_R2R}
R_{st\,R}^{\left(2\right)} = \left(\frac{e \left[\bm{\alpha}_R \times \bm{E}\right]}{\hbar \omega}\right)^2 2 \sqrt{\frac{2 m_e}{\hbar^2}} \frac{\Delta \frac{\lambda_s}{L}\left(1 - e^{-\frac{L}{\lambda_s}}\right)}{\left(\hbar \omega - 2 J\right)^2 + \Delta^2} \frac{J_s}{\left(\varepsilon_F^s\right)^{3/2}} \frac{j_e}{e},
\end{equation}
\begin{equation} \label{Eq3_R2hel}
R_{st\,hel}^{\left(2\right)} = \left(\frac{e \hbar E_z}{2 m_e \omega}\right)^2 2 \sqrt{\frac{2 m_e}{\hbar^2}} q^2 \frac{\Delta \frac{\lambda_s}{L}\left(1 - e^{-\frac{L}{\lambda_s}}\right)}{\left(\hbar \omega - 2 J\right)^2 + \Delta^2} \frac{J_s}{\left(\varepsilon_F^s\right)^{3/2}} \frac{j_e}{e},
\end{equation}
\begin{equation} \label{Eq3_R2J}
R_{st\,J}^{\left(2\right)} = \left(\frac{e \hbar \left(\bm{E} \cdot \frac{\hbar k_F}{J} \frac{\partial J}{\partial \bm{p}}\right)}{m_e \omega}\right)^2 \frac{2}{5} \sqrt{\frac{2 m_e}{\hbar^2}} q^2 \frac{\Delta \frac{\lambda_s}{L}\left(1 - e^{-\frac{L}{\lambda_s}}\right)}{\left(\hbar \omega - 2 J\right)^2 + \Delta^2} \frac{J_s}{\left(\varepsilon_F^s\right)^{1/2} \varepsilon_F} \frac{j_e}{e},
\end{equation}
where we take $\varepsilon_F^s >> \varepsilon_F$ and $J_s >> J$ for simplicity.
The result (\ref{Eq3_R2R})--(\ref{Eq3_R2J}) depends both on the parameters of active region and spin source. Besides, it is proportional to the electric current $j_e$. The total $R_{st}$ is defined as $R_{st} = R_{st}^{\left(1\right)} + R_{st}^{\left(2\right)}$. Since $R_{st}^{\left(1\right)}$ depends only on the parameters of active region it is possible to obtain positive $R_{st}$ by taking approriate spin source and current. For example, if $\varepsilon_F^s >> \varepsilon_F$ the equations (\ref{Eq3_R2R})--(\ref{Eq3_R2J}) are simplified and we have the critical electric current
\begin{equation} \label{Eq3_jcRhel}
j_{e\,R}^c = j_{e\,hel}^c = \frac{\pi}{2} e k_F^3 \frac{\hbar k_F^s}{m_e} \frac{J/\varepsilon_F}{J_s/\varepsilon_F^s} \frac{L/\lambda_s}{1 - e^{-\frac{L}{\lambda_s}}},
\end{equation}
\begin{equation} \label{Eq3_jcJ}
j_{e\,J}^c = \frac{10 \pi}{3} e k_F^3 \frac{\hbar k_F^s}{m_e} \left(\frac{k_F}{k_F^s}\right)^2 \frac{J/\varepsilon_F}{J_s/\varepsilon_F^s} \frac{L/\lambda_s}{1 - e^{-\frac{L}{\lambda_s}}}.
\end{equation}
One can see that the obtained electric current is extremely big for any reasonable system parameters (for example, if we take $\varepsilon_F \sim 1eV, \varepsilon_F^s \sim 5ev, \frac{J}{\varepsilon_F} \sim \frac{J_s}{\varepsilon_F^s} \sim 0.2$ and $L<<\lambda_s$, we obtain $5 \cdot 10^{12} \frac{A}{cm^2}$ from (\ref{Eq3_jcRhel}) and $3 \cdot 10^{12} \frac{A}{cm^2}$ from (\ref{Eq3_jcJ})). The physical meaning of this result is following. There are more electrons in the lower spin subband of the active region than in the upper one. These electrons may transit into the upper subband under the influence of electromagnetic wave if there are free states in the upper subband corresponding to the same electron momentum. This leads to the electromagnetic wave absorption which was discussed for noncollinear magnetic systems in \cite{KarashtinJETP} and is not taken into account directly in papers devoted to electromagnetic wave generation due to transitions between spin subbands in ferromagnet. Roughly, it is needed to inject more electrons into the upper spin subsystem than the excess in the lower one in order to obtain generation of electromagnetic wave. This excess of the electrons is rather big ($\sim N_e \frac{J}{\varepsilon_F} \sim 0.2 N_e$ where $N_e$ is the total electron concentration in the active region). When calculating the emission rate, the electron concentration is multiplied by a weight, i.e. the electron transition probability, which gives (\ref{Eq3_jcRhel}), (\ref{Eq3_jcJ}).

One possible way to overcome this problem is to use a non-metal ferromagnet such as a magnetic semiconductor as an active region. We do not consider such possibility hereafter. Thus, in our consideration the mechanism of electromagnetic wave generation induced by Rashba coupling needs a very big electric current and therefore is not realistic.

Another way that works for non-collinear ferromagnet is to take into account that $\Delta \varepsilon$ depends on the quasimomentum $k_z$ ($z$ is the helicoid axis, see (\ref{Eq1_Bloch})). Indeed, the electrons that participate in electromagnetic wave absorption are mostly concentrated close to the Fermi sphere $\varepsilon \approx \varepsilon_F$ of the active region. On the other hand, the electrons that are injected from the spin source have the energy up to the Fermi energy of spin source $\varepsilon_F^s$. Therefore they may emit electromagnetic wave with different frequency when transit between spin subbands than that is absorbed well (see Figure~{\ref{Fig3}).
\begin{figure}[t]
\center{\includegraphics[width=0.4 \textwidth, keepaspectratio=true]{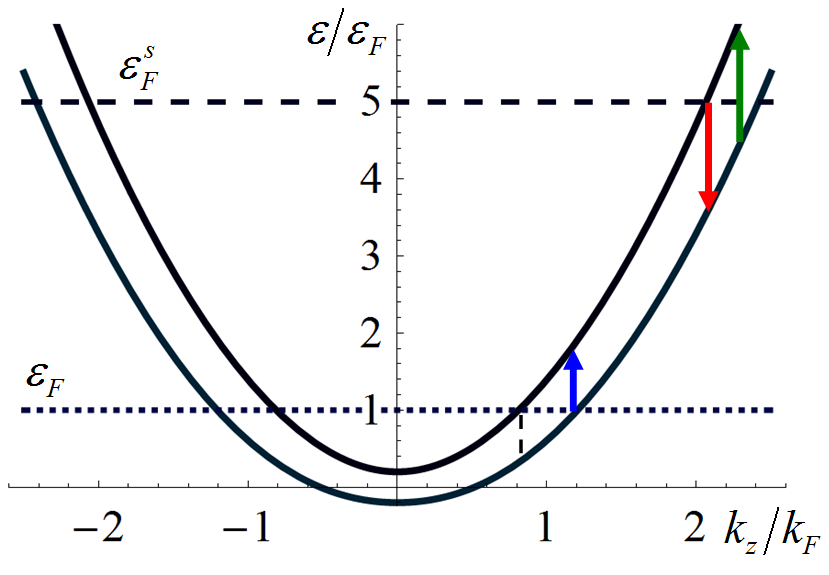}}
\caption{\label{Fig3} (Color online) Spin subbands in a helocoidal ferromagnet for $\varepsilon_F^s = 5 \varepsilon_F, \frac{J}{\varepsilon_F} = 0.2, \frac{q}{k_F} = 0.34$. Three arrows have different length which corresponds to different $\Delta \varepsilon$.}
\end{figure}
Thus, the critical current depends on frequency of electromagnetic wave. For relatively weak spin relaxation processes (small $\Delta$) this current may be small enough to be realized in experiment.  

The exact result that is represented in the form of several integrals may be found in the Appendix. Figure~\ref{Fig4} contains results of numeric calculations of the critical current with respect to dimensionless wave frequency $\Theta = \frac{\hbar \omega}{2 J}$ and dimensionless spin relaxation parameter $\delta = \frac{\Delta}{2 J}$.
\begin{figure}[t]
\center{\includegraphics[width=1 \textwidth, keepaspectratio=true]{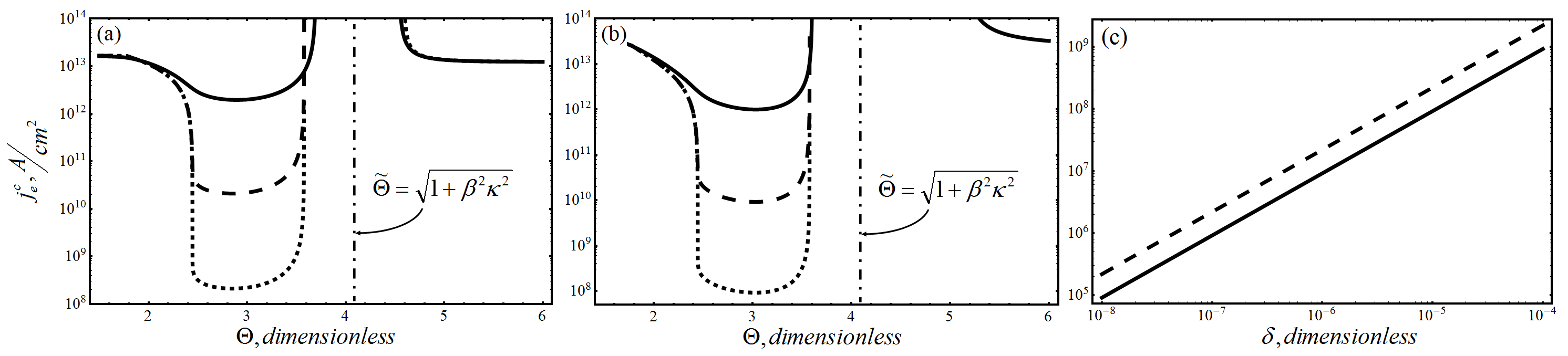}}
\caption{\label{Fig4} Dependence of critical current (a) $j_{e\,hel}^c$ and (b) $j_{e\,J}^c$ on the dimensionless frequency $\Theta = \frac{\hbar \omega}{2 J}$ for the spin relaxation parameter $\delta = \frac{\Delta}{2 J}$ equal to $10^{-1}$ (solid line), $10^{-3}$ (dashed line), and $10^{-5}$ (dotted line). The vertical dash-dotted lines correspond to $\tilde{\Theta} = \sqrt{1 + \beta^2 \kappa^2}$ which is determined by the maximal energy of the injected electrons. (c) Dependence of $j_{e\,J}^c$ (solid line) and $j_{e\,hel}^c$ (dashed line) on $\delta$ at frequency corresponding to $\Theta = 3$.}
\end{figure}
Other parameters are taken as before: $\varepsilon_F = 1 eV, \varepsilon_F^s = 5 eV, \frac{J}{\varepsilon_F} = \frac{J_s}{\varepsilon_F^s} = 0.2, q = 1.8 \cdot 10^7 cm^{-1}$. The spin relaxation length $\lambda_s = 20nm$ which is typical for ferromagnetic metals \cite{Piraux, Bass}, and we take thickness of the active region $L = 5nm$. The dependence of $j_{e \, hel}^c$ and $j_{e \,J}^c$ on frequency is shown in Figure~{\ref{Fig4}a and b, correspondingly, for three different values of $\delta$. It is plotted for $\hbar \omega > 2 J$ because for smaller frequency the critical current is very big, which corresponds to our estimations (\ref{Eq3_jcRhel}), (\ref{Eq3_jcJ}). When the frequency increases (starting at $2 J$) the critical current decreases because the wave absorption by equilibrium electrons (the leftmost (blue) arrow in Figure~\ref{Fig3}) becomes weaker while the wave emission by the injected non-equilibrium electrons is stronger (the middle (red) arrow in Figure~\ref{Fig3}). The wave absorption by electrons which are injected into lower subband starts approximately at the steep decrease of $j_e^c$. This leads to a grow of the cricical current with further increase of the frequency, and finally there is a frequency interval in which $R_{st}$ is negative for any current. This interval corresponds to high $k_z$ at which only electrons injected into lower spin subband exist (the rightmost (green) arrow in Figure~\ref{Fig3}). Contrary to absorption of electromagnetic wave by the equilibrium electrons, the absorption of wave by injected electrons depends on electric current $j_e$ and therefore makes stimulated emission impossible for any electric current in the corresponding frequency interval. At high frequencies, $R_{st}$ becomes positive, but for very big current. The dependence of critical currents for both mechanisms on $\delta$ are plotted in Figure~\ref{Fig4}c for $\frac{\hbar \omega}{2 J} = 3$ which is close to the minimum of $j_e^c$. We see that the critical current decreases linearly and is as small as $10^5 \frac{A}{cm^2}$ for $\delta = 10^{-8}$ which corresponds to spin relaxation time $\tau_s \sim 10^{-6}s$. Although it is hard to imagine such small spin relaxation time, it seems possible to achieve critical current in pulse mode for very clean ferromagnet with $\tau_s \sim 10^{-9}-10^{-10}s$.

\subsection{Radiation power}
Here we perform a simple estimation of the power of stimulated emission of electromagnetic radiation in a helicoid due to two possible mechanisms of electron transitions between spin subbands. Our estimations of critical current and $R_{st}$ in the previous subsection were performed for the stationary non-equilibrium state (\ref{Eq3_eq}) that does not take into account the electromagnetic wave emission. For small deviations of $\delta N$ from $\delta N^*$ determined by (\ref{Eq3_eq}) we may suppose linear dependence of $R_{st}$ on $\delta N$ (this is confirmed by the results of previous work for relatively small current \cite{KarashtinJETPL}). Using equation (\ref{Eq3_G}) we may suppose $G\left(\delta N\right) = G\left(\delta N^*\right) \frac{\delta N}{\delta N^*}$. In order to simplify the calculation, we re-write this equation in terms of $j_e^c$ as
\begin{equation} \label{Eq3_Gn}
G\left(\delta N\right) = \xi \frac{\delta N}{\delta N^*} \left(j_e - j_e^c\right),
\end{equation}
where $\xi$ does not depend on current. Note that we do not discuss the polarization properties of generated waves here, providing just simple estimations.

The equations for the photon density $N_p$ and the non-equilibrium electron density $\delta N$ are written as
\begin{equation} \label{Eq3_Np}
\dot{N}_p = G\left(\delta N\right) N_p - \nu_p N_p,
\end{equation}
\begin{equation} \label{Eq3_dNn}
\dot{\delta N} = \frac{j_e}{e L} \eta \left(1 - e^{-\frac{L}{\lambda_s}}\right) - \frac{\delta N}{\tau_s} - G\left(\delta N\right) N_p,
\end{equation}
where the equation (\ref{Eq3_dN}) for $\delta N$ is modified in order to take into account electron transitions with the emission of electromagnetic wave. Using equation (\ref{Eq3_Gn}) one can easily obtain a stationary state with nonzero photon density:
\begin{equation} \label{Eq3_Npeq}
N_p^{**} = \frac{\delta N^*}{\nu_p \tau_s} \frac{j_e - j_e^{th}}{j_e - j_e^c},
\end{equation}
\begin{equation} \label{Eq3_dNeq}
\delta N^{**} = \delta N^* \frac{\nu_p}{\xi \left(j_e - j_e^c\right)},
\end{equation}
where the threshold current is defined as
\begin{equation} \label{Eq3_jth}
j_e^{th} = j_e^c + \frac{\nu_p \tau_s}{\xi}.
\end{equation}
For the electric current $j_e < j_e^{th}$ there is no stationary state with nonzero photon density, while for $j_e > j_e^{th}$ such stationary state exists hence stimulated electromagnetic wave emission is possible. Obviously, the threshold current $j_e^{th}$ is greater than the critical current $j_e^c$ at which the wave emission exceeds the absorption.
We have taken into account the electromagnetic wave emission by the medium in the form of $R_{st}^{\left(1\right)}$ which does not depend on the electric current. Here we suppose that the photon losses determined by $\nu_p$ are due to the escape of photons from a resonator that contains the active region (i.e. $\nu_p$ is the inverse photon lifetime in the resonator). Therefore the power of electromagnetic radiation emitted from the resonator may be estimated as
\begin{equation} \label{Eq3_Prad}
P_{rad} = \hbar \omega \nu_p N_p^{**} V_{res} = \hbar \omega \frac{\eta j_e}{e L} \left(1 - e^{-\frac{L}{\lambda_s}}\right) \frac{j_e - j_e^{th}}{j_e - j_e^c} V_{res},
\end{equation}
where $V_{res}$ is the resonator volume. It is seen from (\ref{Eq3_jth}) and (\ref{Eq3_Prad}) that the increase of $\nu_p$ leads to grow of $j_e^{th}$ while $P_{rad}$ does not change. Therefore it is necessary to make good resonator in order to obtain stimulated radiation.

In order to estimate the threshold current and radiation power, we take the same parameters as before. The frequency parameter is taken as $\Theta = 3$ which corresponds to the wavelength $\lambda = 1 \mu m$, the spin relaxation time $\tau_s \sim 10^9$. We estimate the resonator volume as $V_{res} \sim \lambda^3$. The inverse photon lifetime in the resonator is taken as $\nu_p \sim 3 \cdot 10^9 s^{-1}$ which corresponds to the quality factor $10^5$ if the resonator length is $\sim \lambda$. For these parameters the critical currents are $j_{e\,hel}^c \approx 2.2 \cdot 10^8 \frac{A}{cm^2}, j_{e\,J}^c \approx 9.2 \cdot 10^7 \frac{A}{cm^2}$, the threshold current is $j_{e\,hel}^{th} \approx 2.4 \cdot 10^8 \frac{A}{cm^2}, j_{e\,J}^{th} \approx 9.9 \cdot 10^7 \frac{A}{cm^2}$. These currents may be acieved in pulse mode. The radiation power for $j_e = 3 \cdot 10^8 \frac{A}{cm^2}$ which is above the threshold for both mechanisms is $P_{rad}^{hel} \approx 78 W, P_{rad}^J \approx 98 W$ inside the pulse. The estimated power is rather big which is caused by high emission rate $R_{st}$ in the considered system. Note that both mechanisms give similar values of threshold currents and radiation power. Therefore they should be taken into account simultaneously.

It should be noted here that our model does not take into account the radiation resistance. Hence very big power is obtained. This is much greater than the power of ohmic losses. In real experiment the increase of the radiation power would decrease current and lead the system to the boundary of generation region. This would keep the radiation power close to the power of ohmic losses by the order of value (several milliwatts). However this deserves a separate study and is beyond the scope of current paper.

\section{\label{Conclusion}Conclusion}
We have theoretically considered possible mechanisms of emission of electromagnetic wave with transition of electrons between spin subbands in a ferromagnet. The probabilities of such electron transitions due to Zeeman coupling caused the magnetic field of the wave, the Rashba spin-orbit coupling in a uniform ferromagnet and also due to two mechanisms (one follows from the minimal coupling, the other is due to the dependence of the exchange constant on the electron momentum) that exist in a non-collinear magnetic system are compared. We have shown that the mechanism of transitions caused by the magnetic field of the wave is much weaker than three other mechanisms, and therefore it could be neglected. We have considered a simple model in which the spin polarized electrons are injected into the ferromagnet by electric current flowing from another ferromagnet with opposite magnetization direction (or magnetization direction at the surface if a non-collinear ferromagnet is considered) through a tunneling barrier. It is shown that the wave absorption in a uniform ferromagnet is very strong which leads to unrealizable critical electric current at which the emission rate becomes positive. In the case of a non-collinear magnetic system, both considered mechanisms give similar estimated threshold currents and radiation power. Such threshold currents may be realized in the experiment. However the considered generation effect is very demanding to the parameters of the system and therefore a very careful optimization of these parameters is necessary.

This work was supported by Russian Science Foundation (Grant No. 19-72-00130).

\appendix
\setcounter{section}{1}
\section*{\label{App}Appendix. The exact expressions for the emission rate}
Here we provide equations for the emission rates and critical current for two mechanisms of wave emission in non-collinear ferromagnet considered in current paper. These equations are exact with respect to $\frac{J}{\varepsilon_F}$ and $\frac{J_s}{\varepsilon_F^s}$. However the probability of electron tunneling through the barrier is the same as in the main body of the paper and is determined by (\ref{Eq3_P}) supposing that the potential barrier height $U >> \varepsilon_F, \varepsilon_F^s$.

In order to describe numerous material parameters of the system, it is convenient to introduce dimensionless parameters $\kappa = \frac{k_F^s}{k_F} = \sqrt{\frac{\varepsilon_F^s}{\varepsilon_F}}, \chi = \frac{J_s}{J}, \beta = \frac{q k_F}{j} (j = \frac{2 m_e}{\hbar^2} J), \delta = \frac{\Delta}{2 J}$, and $\Theta = \frac{\hbar \omega}{2 J}$. Then $R_{st}^{\left(1\right)}$ and $R_{st}^{\left(2\right)}$ are re-written in exact form as
\begin{equation} \label{A_R1hel}
R_{st \, hel}^{\left(1\right)} = - \frac{\pi \hbar k_F^3}{\left(2 J\right)^3} \left(\frac{e \hbar E_z q}{2 m_e}\right)^2 I_{hel}^{\left(1\right)}\left(\Theta\right),
\end{equation}
\begin{equation} \label{A_R2hel}
R_{st \, hel}^{\left(2\right)} = \frac{\pi \hbar k_F^3}{\left(2 J\right)^3} \left(\frac{e \hbar E_z q}{2 m_e}\right)^2 \frac{j_e}{e} \frac{3 \hbar^3}{4 \pi m_e \varepsilon_F^2} \frac{\lambda_s}{L} \left(1 - e^{-\frac{L}{\lambda_s}}\right) \frac{1}{\kappa_s^6 - 1} I_{hel}^{\left(2\right)}\left(\Theta\right),
\end{equation}
\begin{equation} \label{A_R1J}
R_{st \, J}^{\left(1\right)} = - \frac{\pi \hbar k_F^3}{\left(2 J\right)^3} \left(\frac{e \hbar q}{m_e} \left(\bm{E} \cdot \frac{\hbar k_F}{J} \frac{\partial J}{\partial \bm{p}}\right)\right)^2 I_J^{\left(1\right)}\left(\Theta\right),
\end{equation}
\begin{equation} \label{A_R2J}
R_{st \, J}^{\left(2\right)} = \frac{\pi \hbar k_F^3}{\left(2 J\right)^3} \left(\frac{e \hbar q}{m_e} \left(\bm{E} \cdot \frac{\hbar k_F}{J} \frac{\partial J}{\partial \bm{p}}\right)\right)^2 \frac{j_e}{e} \frac{3 \hbar^3}{4 \pi m_e \varepsilon_F^2} \frac{\lambda_s}{L} \left(1 - e^{-\frac{L}{\lambda_s}}\right) \frac{1}{\kappa_s^6 - 1} I_J^{\left(2\right)}\left(\Theta\right),
\end{equation}
where the dimensionless frequency functions $I_{hel}^{\left(1,2\right)}\left(\Theta\right), I_J^{\left(1,2\right)}\left(\Theta\right)$ are defined in the form of integrals:
\begin{eqnarray} \label{A_I1}
I_{\{hel,J\}}^{\left(1\right)}\left(\Theta\right) = \int_0^{x_-}{dx F_{\{hel,J\}}\left(x,\Theta\right) \left(1 - x^2 + \gamma \sqrt{1 + \beta^2 x^2}\right)} \\ \nonumber
- \int_0^{x_+}{dx F_{\{hel,J\}}\left(x,\Theta\right) \left(1 - x^2 - \gamma \sqrt{1 + \beta^2 x^2}\right)},
\end{eqnarray}
\begin{eqnarray} \label{A_I2} \nonumber
I_{\{hel,J\}}^{\left(2\right)}\left(\Theta\right) = \int_0^{x_+}{dx F_{\{hel,J\}}\left(x,\Theta\right) \left(\kappa^2 - 1\right) \left(x^2 + \gamma \left(\sqrt{1 + \beta^2 x^2} + \chi\right)\right)} \\ \nonumber
+ \int_{x_+}^{x_+^s}{dx F_{\{hel,J\}}\left(x,\Theta\right)\left(\kappa^2 - x^2 - \gamma \sqrt{1 + \beta^2 x^2}\right) \left(x^2 + \gamma \left(\sqrt{1 + \beta^2 x^2} + \chi\right)\right)} \\
- \int_{\tilde{x}}^{x_-^a}{dx F_{\{hel,J\}}\left(x,\Theta\right) \left(\kappa^2 - \mu\right) \left(x^2 - \gamma \left(\sqrt{1 + \beta^2 x^2} + \chi\right)\right)} \\ \nonumber
- \int_{x_-^a}^{x_-^s}{dx F_{\{hel,J\}}\left(x,\Theta\right)\left(\kappa^2 - x^2 + \gamma \sqrt{1 + \beta^2 x^2}\right) \left(x^2 - \gamma \left(\sqrt{1 + \beta^2 x^2} + \chi\right)\right)},
\end{eqnarray}
\begin{eqnarray} \label{A_Fhel}
F_{hel}\left(x,\Theta\right) = \frac{\delta}{\Theta^2 \left(1 + \beta^2 x^2\right) \left(\delta^2 + \left(\Theta - \sqrt{1 + \beta^2 x^2}\right)^2\right)}, \\ \label{A_FJ}
F_{J}\left(x,\Theta\right) = x^2 F_{hel}\left(x,\Theta\right),
\end{eqnarray}
the constant $\mu$ in (\ref{A_I2}) is defined as $\mu = \max{\left(1, \gamma \chi\right)}$ depending on system parameters, and the dimensionless integration limits are the following:
\begin{equation} \label{A_x}
x_\pm = \sqrt{1 + \frac{\beta^2 \gamma^2}{2} \mp \sqrt{1 + \beta^2 + \frac{\beta^4 \gamma^2}{4}}},
\end{equation}
\begin{equation} \label{A_xs}
x_\pm^s = \sqrt{\kappa^2 + \frac{\beta^2 \gamma^2}{2} \mp \sqrt{1 + \kappa^2 \beta^2 + \frac{\beta^4 \gamma^2}{4}}},
\end{equation}
\begin{equation} \label{A_xt}
\tilde{x} = \sqrt{\gamma \chi + \frac{\beta^2 \gamma^2}{2} + \gamma \sqrt{1 + \beta^2 \gamma \chi + \frac{\beta^4 \gamma^2}{4}}},
\end{equation}
\begin{equation} \label{A_xa}
x_-^a = \max{\left(x_-, \tilde{x}\right)}.
\end{equation}
Note that $\tilde{x}$ stands for the smallest $k_z$ at which the electrons are injected into the lower spin subband of the active region from the upper spin subband of the spin source (in the latter, there are no electrons with the energy corresponding to small $k_z$ in the lower subband in the active region; $k_z$ is not conserved at the boundary). The third integral in the right-hand part of (\ref{A_I2}) is zero if $\tilde{x} > x_-$. This condition depends on the system parameters in a complex way. But is exactly true if $\varepsilon_F < J_s$ ($\gamma \chi > 1$ in dimendionless notation).
In (\ref{A_I2}), we take into account both reduction of the electromagnetic wave absorption by equilibrium electrons due to the appearance of electrons in the upper spin subband lead by the electric current and absorption of electromagnetic wave by the electrons injected into lower spin subband which is stronger than emission due to the inverse process at high $k_z$ for which there are no injected electrons in the upper spin subband.

Applying a condition of positive $R_{st}$ leads us to the critical current $j_e^c\left(\theta\right)$
\begin{equation} \label{A_jc}
j_{e\,\{hel,J\}}^c \left(\theta\right) = e \frac{\pi}{3} k_F^3 \frac{\hbar k_F}{m_e} \frac{L/\lambda_s}{1 - e^{-\frac{L}{\lambda_s}}} \left(\kappa_s^6 - 1\right) \frac{I_{\{hel,J\}}^{\left(1\right)}\left(\Theta\right)}{I_{\{hel,J\}}^{\left(2\right)}\left(\Theta\right)}.
\end{equation}
One can see that the exact equation (\ref{A_jc}) for the critical current contains dependence on the wave frequency which is very important if small parameter $\delta$ (that stands for the spin relaxation) is taken.

\section*{Reference}
\bibliography{gen}

\end{document}